\def\a{\alpha}\def\b{\beta}

\def\de{\partial}
\def\inf{\infty}\def\id{\equiv}\def\ha{{1\over 2}}

\def\section#1{\bigskip\noindent{\bf#1}\smallskip}
\def\nota{\footnote{$^\dagger$}}

\def\PR#1{Phys.\ Rev.\ {\bf#1}}\def\CQG#1{Class.\ Quantum Grav.\ {\bf#1}} 
\def\NP#1{Nucl.\ Phys.\ {\bf#1}} 
 
\def\NC#1{Nuovo Cimento {\bf#1}}

\def\ref#1{\medskip\everypar={\hangindent 2\parindent}#1}
\def\beginref{\begingroup
\bigskip
\centerline{\bf References}
\nobreak\noindent}
\def\endref{\par\endgroup}

\magnification=1200

{\nopagenumbers
\line{April 2000\hfil INFNCA-TH0007}
\vskip80pt
\centerline{\bf A note on the infinite-dimensional symmetries}
\centerline{\bf  of classical hamiltonian systems}
\vskip40pt
\centerline{{\bf S. Mignemi}\nota{e-mail:MIGNEMI@CA.INFN.IT}}
\vskip10pt
\centerline {Dipartimento di Matematica, Universit\`a di Cagliari}
\centerline{viale Merello 92, 09123 Cagliari, Italy}
\centerline{and INFN, Sezione di Cagliari} 

\vskip80pt
{\noindent 
We show that any Hamiltonian system with one degree of freedom
is invariant under a $w_\infty$ algebra of symmetries.}
\vfil\eject}

Consider a hamiltonian system with one degree of freedom and time-independent
hamiltonian
$$H=H(p,q).$$
It is well known that $H(p,q)$ is a constant of motion, $H=\a$.
Another constant of motion $\b$ can be obtained by solving the Hamilton-Jacobi
equation
$$H\left(q,{dW\over dq}\right)=\alpha,$$
from which one can define a new coordinate $Q={\de W\over\de\a}=t+\b$.
One has
$$\b=Q(q,\a)-t\id G(p,q),$$
and Poisson brackets
$$\{G(p,q),H(p,q)\}=1.$$
Moreover,
$${dG\over dt}={\de G\over\de t}+\{G,H\}=0.$$

Consider now the quantities
$$w_{m,n}=H^{m+1}G^n.\eqno(1)$$
Being algebraic functions of conserved quantities they are of course 
conserved,
${dw\over dt}=0$. Moreover, their Poisson brackets obey the $w_\inf$ algebra
$$\{w_{m,n},w_{m',n'}\}=[m(n'+1)-m'(n+1)]w_{m+m',n+n'}.\eqno(2)$$
By construction, the constants of motion $w_{m,n}$ are 
functionally dependent on $H$ and $G$. 
Nevertheless, they generate a non-trivial algebra of symmetries.
Of course, this unusual situation is due to the explicit
dependence of $G$ on time.

The simplest example is given by the harmonic oscillator, with hamiltonian
$$H=\ha\left(p^2+q^2\right),$$
in which case,
$$G=-t-\arccos{q\over\sqrt{2H}}.$$
A less trivial example is given by conformal mechanics [1,2] and is discussed in 
detail in [3]. In this case,
$$H=\ha{p^2\over f(pq)},\qquad G=-t+{pq\over\sqrt{2H}}.$$

The existence of infinite-dimensional symmetries in systems with one degree
of freedom was first noticed in the case of the inverted harmonic oscillator
[4] and of conformal mechanics [2] as an invariance of the volume of the
phase space, but no detailed account of the realization of the symmetry
in terms of conserved charges or of its physical significance was given.
Only in ref. [3] the special case of conformal mechanics was discussed from
this point of view.

The $w_\inf$ symmetries discussed above can be easily generalized to the
case of integrable systems with a finite number of degrees of freedom.
While in classical mechanics this invariance acts at a formal level, 
its extension to  quantum systems may yield non-trivial consequences,
because of the possible appearance of central extensions of the algebra.
This fact would be especially interesting in the case of conformal
quantum mechanics, for its implications on the $AdS_2/CFT_1$ duality [2,5].

\beginref
\ref[1] V. de Alfaro, S. Fubini and G. Furlan, \NC{34A}, 569 (1976).
\ref[2] S. Cacciatori, D. Klemm and D. Zanon, \CQG{17}, 1731 (2000).
\ref[3] M. Cadoni, P. Carta and S. Mignemi, "A realization of the 
infinite-dimensional symmetries of conformal mechanics",
{\tt hep-th/0004107}.
\ref[4] E. Witten, \NP{B373}, 187 (1982).
\ref[5] M. Cadoni and S. Mignemi, \PR{D59}, 081501 (1999);
\NP{B557}, 165 (1999).
\endref
\end